\documentstyle[12pt]{article}
\begin{document}

\def\be{\begin{equation}}
\def\ee{\end{equation}}
\def\bea{\begin{eqnarray}}
\def\eea{\end{eqnarray}}
\def\nn{\nonumber \\}
\def\e{{\rm e}}
\def\tr{{\rm tr}}

\begin{flushright} 
NDA-FP-28\ \ \ \ \ \ \ \\
September 1996
\end{flushright}

\vfill

\begin{center}
{\large\bf Casimir effect in the curved background \\
 and the black hole in three dimensions}

\vfill

{\normalsize
Shin'ichi NOJIRI\footnote{nojiri@cc.nda.ac.jp}}

\vfill

{\it
Department of Mathematics and Physics

National Defence Academy, Yokosuka, 239, JAPAN}

\end{center}

\vfill

\begin{abstract}
We consider the quantum correction to the Lagrangean 
by the massless free boson 
in the curved background in three dimensions where 
one of the coordinates is periodic. 
The correction term is given by an expansion of the metric 
with respect to the derivative and the first term 
expresses to the usual Casimir energy.
As an application, we investigate the change of the 
geometry in three dimensional black hole due to the 
quantum effect and 
we show that the geometry becomes like that of the 
Reissner-Nordstr\o m solution. 
\end{abstract}

\newpage

Since the finite size effects, {\it e.g.} the Casimir 
effect, in the curved space were 
usually calculated by fixing the background geometry 
\cite{eorbz}, it is not always useful to consider 
the quantum back reaction to the geometry.
In this paper, we consider the quantum correction 
to the Lagrangean by the massless free boson 
in the curved background in three dimensions where 
one of the coordinates is periodic. 
The correction term is given by an expansion of the metric 
with respect to the derivative and the first term 
express the usual Casimir energy.
As an application, we investigate the change 
of the geometry in three dimensional black hole 
due to the quantum effect and 
we show that the geometry becomes like that of the 
Reissner-Nordstr\o m solution. 

The d'Alembertian $\Box$ for the free massless boson 
$\varphi$ is defined by
\be
\label{box}
\Box\equiv \partial_\mu \sqrt{g} g^{\mu\nu} 
\partial_\nu 
\ee
and the general coordinate transformation invariant 
measure for the boson $\varphi$ is given by
\be
\label{measure0}
\Delta\varphi^2=\int dx^3 \sqrt{g} \delta \varphi(x)^2 \ .
\ee
If we define a new field $\widetilde \varphi$ by
\be
\label{new}
\widetilde\varphi = g^{1 \over 4}\varphi\ ,
\ee
the measure becomes trivial: 
\be
\label{measure1}
\Delta\widetilde\varphi^2=\int dx^3 \delta \widetilde\varphi(x)^2 \ .
\ee
Then the d'Alembertian $\widetilde\Box$ 
for $\widetilde\varphi$ is 
given by
\be
\label{box2}
\widetilde\Box\equiv g^{-{1 \over 4}}
\partial_\mu \sqrt{g} g^{\mu\nu} 
\partial_\nu g^{-{1 \over 4}}\ .
\ee
We now evaluate $-{1 \over 2}\tr \ln \Box $ by the 
heat kernel method as follows
\be
\label{heat}
-{1 \over 2}\tr \ln \Box 
\equiv -{1 \over 2}\lim_{\epsilon\rightarrow 0}
\int_{\epsilon^2}^\infty {dt \over t}
\widetilde \tr\e^{t\widetilde\Box}\ .
\ee
Here $\widetilde\tr$ is defined by 
\bea
\label{tr}
\widetilde\tr {\cal O} &\equiv&
\int d^3x \int_{-\infty}^\infty {dk_1 \over 2\pi} 
\int_{-\infty}^\infty {dk_2 \over 2\pi} 
\sum_{n=-\infty}^\infty {1 \over 2\pi} \nn 
&\times& \e^{-i\left(k_1x^1+k_2x^2+ n x^3
\right)} {\cal O}
\e^{i\left(k_1x^1+k_2x^2+ n x^3\right)}\ ,
\eea
and we assume that $x^3$ is a coordinate with the period 
$2\pi$:
\be
\label{peri}
x^3 \sim x^3 + 2\pi\ .
\ee 
When we expand $\tr \ln \Box $ with respect to the 
derivatives, we find the following expression by 
replacing $t$ with $\epsilon^2 t$; 
\bea
\label{heat2}
\tr \ln \Box &=& \int_1^\infty 
{dt \over (2\pi)^2\epsilon^2 t^2}
\sum_{n=-\infty}^\infty 
\Bigl\{ f^{(0)}+t\epsilon^2\sum_{m=0} f_m^{(1)}{d^m \over d\alpha^m} \nn
+ \cdots 
&& +(t\epsilon^2)^l\sum_{m=0} f_m^{(k)}{d^m \over d\alpha^m}+\cdots
\Bigr\}\e^{-\alpha t \epsilon^2 n^2}
\eea
Here $f_m^{(k)}$ contains $2k$-derivatives of 
the metric $g_{\mu\nu}$ and 
\bea
\label{alpha}
\alpha&=&{1 \over g_{33}} \\
\label{f0}
f^{(0)}&=&\sqrt{g \over g_{33}} \ .
\eea

We now evaluate the first term $f^{(0)}$,
\bea
\label{term0}
&&\sqrt{g \over g_{33}} \int_1^\infty {dt \over (2\pi)^2 
\epsilon^2 t^2}
\sum_{n=-\infty}^\infty 
\e^{-{\epsilon^2 \over g_{33}}n^2 t
} \nn
&& = {\sqrt{g} \over  (2\pi)^2 \epsilon^3}
\int_1^\infty {dt \over t^{5 \over 2}}
\left(1+2\sum_{n=1}^\infty 
\e^{-{\pi^2 g_{33} \over \epsilon^2 t}n^2 }\right) \ .
\eea
Here we have used the modular transformation 
\be
\label{modu}
1+2\sum_{n=1}^\infty 
\e^{-a\pi n^2}=
\alpha^{-{1 \over 2}}\left(
1+2\sum_{n=1}^\infty 
\e^{-{\pi \over a}n^2}\right)\ .
\ee
The first divergent term can be absorbed into the renormalization 
of the cosmological constant. We rewrite the remaining 
term as follows,
\bea
\label{rem}
&& {2\sqrt{g} \over  (2\pi)^2 \epsilon^3}
\int_1^\infty {dt \over t^{5 \over 2}}
\sum_{n=1}^\infty 
\e^{-{\pi^2 g_{33} \over \epsilon^2 t}n^2 } \nn
&& = {2\sqrt{g} \over  (2\pi)^2\pi^3 
(g_{33})^{3 \over 2}}
\sum_{n=1}^\infty {1 \over n^3}
\int_0^{\pi^2 g_{33} n^2 \over \epsilon^2} ds s^{1 \over 2}\e^{-s} 
\hskip 1cm \left(
s\equiv {\pi^2 g_{33} n^2 \over \epsilon^2 t}\right) \nn
&& \stackrel{\epsilon \rightarrow 0}{\rightarrow}
{2\sqrt{g} \over  (2\pi)^2\pi^3 (g_{33})^{3 \over 2}}
\Gamma({3\over 2})\zeta(3) \nn
&& ={\sqrt{g} \over (2\pi)^2\pi^{5 \over 2} 
(g_{33})^{3 \over 2}}\zeta(3) 
\eea

The second term in Eq.(\ref{heat2}) 
\be
\label{term1}
\sum_{m=0} f_m^{(1)}{d^m \over d\alpha^m}
\int_1^\infty {dt \over (2\pi)^2 t}
\e^{-\alpha t \epsilon^2 n^2}
\ee
can be also evaluated by 
using the modular transformation (\ref{modu}) 
as follows,
\bea
\label{heat3}
&&\int_1^\infty 
{dt \over (2\pi)^2 t}
\sum_{n=-\infty}^\infty
\e^{-\alpha t \epsilon^2 n^2} \nn
&&=\int_1^\infty 
{dt \over (2\pi)^2 \epsilon \alpha^{1 \over 2} 
t^{3 \over 2}}
\left(1+2\sum_{n=1}^\infty
\e^{-{\pi^2 \over \alpha t \epsilon^2} n^2} \right)
\eea
The first divergent term can be also absorbed into the renormalization 
of the gravitational constant and the 
remaining term can be rewritten as follows, 
\bea
\label{second}
&&2 \int_1^\infty 
{dt \over (2\pi)^2 \epsilon \alpha^{1 \over 2} 
t^{3 \over 2}}
2\sum_{n=1}^\infty
\e^{-{\pi^2 \over \alpha t \epsilon^2} n^2} \nn
&& ={2 \over (2\pi)^2 \pi }
\sum_{n=1}^\infty {1 \over n}
\int_0^{\pi^2 n^2 \over \alpha \epsilon^2} 
ds s^{-{1 \over 2}}\e^{-s} 
\hskip 1cm \left(
s\equiv {\pi^2 n^2 \over \alpha \epsilon^2 t}\right) \nn
&& \stackrel{\epsilon \rightarrow 0}{\rightarrow}
{2 \over (2\pi)^2 \pi^{1 \over 2} }\zeta(1)
\eea
Note that the result is $\alpha$-independent.
Therefore $m\neq 0$ terms in Eq.(\ref{term1}) do
not contribute. $f^{(1)}_0$ in the remaining term 
has the following form,
\bea
\label{f10}
f^{(1)}_0&=& \sqrt{g \over g_{33}}\Bigl\{
g_{ij}g_{kl}g_{mn}\Bigl(
-{1 \over 8}g^{i\mu}\partial_\mu g^{km}g^{j\nu}
\partial_\nu g^{ln}
-{1 \over 16}g^{i\mu}\partial_\mu g^{kl}g^{j\nu}
\partial_\nu g^{mn} \nn
&& -{1 \over 4}g^{i\mu}\partial_\mu g^{lm}g^{k\nu}
\partial_\nu g^{jn} 
-{1 \over 4}g^{i\mu}\partial_\mu g^{kl}g^{m\nu}
\partial_\nu g^{jn} 
-{1 \over 4}g^{i\mu}\partial_\mu g^{jm}g^{k\nu}
\partial_\nu g^{ln}\Bigr) \nn
&& + g_{ij}g_{kl}\Bigl(
{1 \over 12}g^{\mu\nu}\partial_\mu g^{ij}
\partial_\nu g^{kl}
+{1 \over 6}g^{\mu i}\partial_\mu g^{\nu j}
\partial_\nu g^{kl}
+{1 \over 6}g^{\mu i}g^{\nu j}\partial_\mu 
\partial_\nu g^{kl} \nn
&&+{1 \over 4}g^{\mu i}\partial_\nu g^{\nu j}
\partial_\mu g^{kl} 
+{1 \over 12}g^{\mu\nu}\partial_\mu g^{ik}
\partial_\nu g^{jl}
+{1 \over 6}g^{\mu i}\partial_\mu g^{\nu k}
\partial_\nu g^{lj} \nn
&& +{1 \over 6}g^{\mu i}g^{\nu k}
\partial_\mu \partial_\nu g^{jl}
+{1 \over 4}g^{\mu i}\partial_\nu g^{\nu k}
\partial_\mu g^{lj}\Bigr) \nn
&&+g_{ij}\Bigl(-{1 \over 4}g^{\mu\nu}
\partial_\mu \partial_\nu g^{ij}
-{1 \over 4}\partial_\mu g^{\mu\nu}\partial_\nu g^{ij}
-{1 \over 2}g^{\mu i}\partial_\mu \partial_\nu g^{\nu j}
-{1 \over 4}\partial_\mu g^{\mu i}
\partial_\nu g^{\nu j} \Bigr)\nn
&&-{1 \over 16}g^{\mu\nu}g_{\rho\sigma}
\partial_\nu g^{\rho\sigma} g_{\eta\zeta}
\partial_\nu g^{\eta\zeta}
+{1 \over 4}g^{\mu\nu}\partial_\mu g_{\rho\sigma}
\partial_\nu g^{\rho\sigma}
+{1 \over 4}g^{\mu\nu}\partial_\mu 
\partial_\nu g^{\rho\sigma}g_{\rho\sigma} \nn
&& +{1 \over 4}\partial_\mu g^{\mu\nu} 
\partial_\nu g^{\rho\sigma}g_{\rho\sigma}\Bigr\}\ .
\eea
Here the roman indeces $i,j,k,\cdots =1,2$ and the 
Greek ones $\mu,\nu,\rho,\cdots=1,2,3$. 
Then the finite part of $-{1 \over 2}\tr \ln \Box $ 
has the following form
\bea
\label{trln}
&&-{1 \over 2}\tr \ln \Box |_{\mbox{{\small finite part}}} 
\nn
& = c_0 {\sqrt{g} \over (g_{33})^{3 \over 2}}
+  c_1f^{(1)}_0 + \mbox{higher derivative term}
\eea
Here
\be
\label{c0c1}
c_0 =-{\zeta(3) \over 2(2\pi)^2\pi^{5 \over 2}} \ , 
\hskip 1cm
c_1=-{\zeta(1) \over (2\pi)^2 \pi^{1 \over 2} }\ .
\ee
Note that the parameter $c_0$ is negative 
but if we consider the contribution from the free 
fermions\footnote{
The quantum back reaction of Nambu-Jona-Lasino type 
fermion was investigated in Ref.\cite{mos}.} 
instead of the free boson, $c_0$ can be 
positive.

The first term in Eq.(\ref{trln}) expresses 
the Casimir energy in three dimensions.
When the three dimensional space-time is given by 
the direct product of a circle ($S^1$) and 
two dimensional space ($R^2$), $g_{33}$ is given by 
\be
\label{s1r2}
g_{33}=\sqrt{L}\ .
\ee
Here $L$ is the radius of $S^1$. 
If we substitute Eq.(\ref{s1r2}) into Eq.(\ref{rem}), 
the first term in Eq.(\ref{trln}) is proportional to 
${1 \over L^3}\sqrt g$ and the second term to
${1 \over L}\sqrt g R$ and reproduce the known results 
\cite{eorbz}.
The full expression for $-{1 \over 2}\tr \ln \Box $ 
would contain an infinite number of derivatives and 
would be non-local\footnote{
Such a non-local effective action was obtained by using 
renormalization group techniqes in Ref.\cite{eo}.} 
since $-{1 \over 2}\tr \ln \Box $ 
should depend on the global information like periodic 
dimensions, {\it e.g.} $L$ in Eq.(\ref{s1r2}).
The expression in Eq.(\ref{trln}) does not appear 
to be covariant under the general coordinate 
transformation.
The covariance would be, however, restored in a non-local 
way if we could obtain the full expression since the 
heat kernel method used here manifestly preserves the 
general coordinate invariance.

We now consider the change of the geometry in the three
dimensional black hole \cite{btz} due to the 
quantum correction obtained in (\ref{trln}).

The geometry of the three dimensional 
static (non-rotating) black hole is 
obtained by imposing a periodic boundary condition 
on the anti-de Sitter space, whose metric has 
the following form
\be
\label{AdS}
ds^2=l^2\left({dz^2+d\beta^2-d\gamma^2 \over z^2}\right) \ .
\ee
Here $\{z,\beta,\gamma\}$ are Poincar\'e coodinates 
and $\Lambda=-l^{-2}$ is the cosmological constant.
The periodic boundary condition is given by 
the following identification,
\be
\label{period}
(z,\beta,\gamma)\sim\e^{{r_0 \over l}p}(z,\beta,\gamma)\ .
\ee
Here $p=0,\pm 2\pi,\pm4\pi,\cdots$.
If we define new coordinates $\{ r,\phi,t\}$ by
\bea
\label{coord1}
r>r_0&& \\
z&=&{r_0 \over r}\e^{-{r_0 \over l}\phi} \nn
\beta &=& \left(1-{r_0^2 \over r^2}\right)^{1 \over 2}
\e^{-{r_0 \over l}\phi} \cosh {r_0 t \over l^2} \nn
\gamma &=& -\left(1-{r_0^2 \over r^2}\right)^{1 \over 2}
\e^{-{r_0 \over l}\phi} \sinh {r_0 t \over l^2} \ , \\
r<r_0&& \nn
z&=&{r_0 \over r}\e^{-{r_0 \over l}\phi} \nn
\beta &=& -\left({r_0^2 \over r^2}-1\right)^{1 \over 2}
\e^{-{r_0 \over l}\phi} \cosh {r_0 t \over l^2} \nn
\gamma &=& \left({r_0^2 \over r^2}-1\right)^{1 \over 2}
\e^{-{r_0 \over l}\phi} \sinh {r_0 t \over l^2} \ , 
\eea
the metric in Eq.(\ref{AdS}) is rewritten by,
\bea
\label{bh}
ds^2&=&-\left(-M+{r^2 \over l^2}\right) dt^2
+\left(-M+{r^2 \over l^2}\right)^{-1} dr^2
+r^2d\phi^2
\nn
&\equiv&g_{\mu\nu}^{(0)}dx^\mu dx^\nu\ .
\eea
Here the mass $M$ is defined by
\be
\label{mass}
M={r_0^2 \over l^2} \ .
\ee
The periodic boundary condition (\ref{period}) is given 
in terms of these coordinates by
\be
\label{period2}
\phi \sim \phi + 2\pi 
\ee

In order to evaluate $ -{1 \over 2}\tr \ln \Box $, 
we Wick-rotate the time coordinate $t$:
\be
\label{wick}
t=i\tau \ .
\ee
If we define a new radius coordinate $\rho$ by
\be
\label{rho}
r=r_0\cosh{\rho \over l}
\ee
the metric has the following form
\be
\label{bh2}
ds^2=\left({r_0 \over l}\right)^2\sinh^2{\rho \over l} 
d\tau^2+d\rho^2
+r_0^2\cosh^2{\rho \over l}d\phi^2\ .
\ee
In order to avoid the conical singularity at $\rho=0$, 
the Euclid time $\tau$ should be a peiodic coordinate 
with the period ${2\pi l^2 \over r^0}$, which give the 
black hole temperature $T$:
\be
\label{temp}
kT={r_0 \over 2\pi l^2}
\ee
($k$ is the Boltzman constant).
By further defining new coordinates $(x^1,x^2,x^3)$ by
\bea
\label{coordx}
x^1&=&\rho\cos {r_0 \tau \over l^2} \nn
x^2&=&\rho\sin {r_0 \tau \over l^2} \\
x^3&=&\phi,
\eea
we find that the Wick-rotated space-time 
is topologically $S^1\times R^2$ (Of course 
this does not mean that the Wick-rotated 
space-time is 
not the direct product of $S^1$ and $R^2$.). 
Here the coordinate of $S^1$ is given 
by $x^3=\phi$ and those of $R^2$ by $(x^1,x^2)$.

In the following, we solve the effective equations
of motion,
\be
\label{efeqm}
0=R_{\mu\nu}-{1 \over 2}g_{\mu\nu}(R+2l^{-2})+
c_0\left( -{1 \over 2}{g_{\mu\nu} \over 
(g_{33})^{3 \over 2}}
+{3 \over 2}{\delta_\mu^3 \delta_\nu^3 \over 
(g_{33})^{1 \over 2}}\right)
\ee
which is obtained from the effective Lagrangean including 
the quantum correction (\ref{trln}) when we neglect  
the terms containing the derivatives.
By multiplying $g^{\mu\nu}$, we find 
the scalar curvature $R$ is given by  
\be
\label{curv}
R=-6l^{-2}\ .
\ee
This tells that 
the derivative for the metric is $O(l^{-1})$. 
Since the $k$-th term in Eq.(\ref{trln}) contains 
$2k$-derivatives, the contribution from the $k$-th term 
is $O(l^{-2k})$.
Therefore we can naturally neglect the derivative 
terms which contains $f_m^{(k)}$ ($k\geq 1$) 
in (\ref{trln}) when $l^{-2}$ 
(minus the cosmological constant) is small.

In three dimensions, the three gauge degrees of freedom 
of the general coordinate transformation 
can be fixed by choosing the condition,
\be
\label{gaugecon}
g_{\mu\nu}=0\ , \hskip 1cm \mu\neq\nu\ .
\ee
The residual symmetry is given by
\be
\label{residual}
x^\mu \rightarrow f^\mu (x^\mu) \ .
\ee
Here $f^\mu$ does not depend on $x^\nu$ when $\mu\neq\nu$.
If we assume  the rotational invariance, we can also 
choose a coordinate 
system where the metric $g_{\mu\nu}$ does not depend on $\phi$:
\be
\label{phi}
g_{\mu\nu}=g_{\mu\nu}(r,t) \ .
\ee
The radial coordinate $r$ is defined by
\be
\label{radial}
g_{\phi\phi}=r^2\ .
\ee
If we further restrict the form of the metric as follows\footnote{
This gives the general solution when the system has the rotational
invariance.}
\be
\label{metric}
g_{tt}=-f(r)\ , \ \ g_{rr}={1 \over f(r)}\ , \ \ g_{\phi\phi}=r^2\ ,
\ \ \mbox{others}=0\ ,
\ee
the Ricci curvatures $R_{\mu\nu}$ are given by
\bea
\label{ricci}
R_{tt}&=&{1 \over 2}f(r)f''(r)+{1 \over 2r}f(r)f'(r) \nn
R_{rr}&=&-{1 \over 2f(r)}f''(r)-{1 \over 2rf(r)}f'(r) \nn
R_{\phi\phi}&=&-rf'(r) \nn
\mbox{others}&=&0
\eea
Then the solution of Eq.(\ref{efeqm}) is given by
\be
\label{sol}
f=-M+\left({r \over l}\right)^2-{c_0 \over r}
\ee
Note that the metric $g_{\mu\nu}^{(0)}$ 
in Eq.(\ref{bh}) is reproduced 
in the limit of $c_0\rightarrow 0$.

The geometry given by the metric (\ref{metric}) and (\ref{sol}) 
is crucially depend on the sign of $c_0$. 
Since 
\be
f'(r)={2r \over l^2}+{c_0 \over r^2}
\ee
\begin{itemize}
\item When $c_0>0$ (free fermion), $f'(r)>0$. 
Therefore there is only one horizon.
\item When $c_0<0$ (free boson), $f'(r)=0$ if 
$r=\tilde r_0=\left(-{c_0 l^2 \over 2}
\right)^{1 \over 3}$. Since 
\be
\label{fr0}
f(\tilde r_0)=-M+M_0\ , \ \ M_0
\equiv {3 \over 2}\left({2c_0^2 \over l^2}
\right)^{1 \over 3}>0
\ee
\begin{itemize}
\item When $M>M_0$, there are two horizons.
\item When $M<M_0$, there are no horizons.
\item When $M=M_0$, there is one horizon.
\end{itemize}
\end{itemize}
Here the horizon is defined by $f(r)=0$. When $c_0<0$, the 
behavior of the solution is very similar to that of the Reissner-Nordstr\o m solution. 
Especially the solution of $M=M_0$ corresponds to the 
extreme black hole solution. In the extreme solution, 
the gravity balances with the energy of electro-magnetic 
force. 
In our case, the gravity would balance with 
the Casimir energy.
If the black hole which has the mass $M$ greater than 
$M_0$ could lose its mass by the Hawking radiation, the 
radiation would stop when the mass $M$ becomes equal 
to $M_0$ and the extreme black hole would remain as a  
remnant.

\ 

\noindent
{\bf Acknowledgments}

The authour gratefully acknowledges 
A. Sugamoto and S.D. Odintsov for the discussions.

\end{document}